\documentclass[fleqn,twoside]{article}
\usepackage{espcrc2}

\usepackage{graphicx}
\usepackage{epsfig}
 \usepackage[figuresright]{rotating}

\newcommand{\AmS}{{\protect\the\textfont2
  A\kern-.1667em\lower.5ex\hbox{M}\kern-.125emS}}

\title{QCD spectroscopy with three light quarks}

\author{F. Farchioni\address[DESY]{Deutsches Elektronen-Synchrotron, DESY,
        Notkestr. 85, D-22603 Hamburg, Germany}%
        \thanks{Address after October 1st:
        Institut f\"ur Theoretische Physik, Universit\"at M\"unster, 
        Wilhelm-Klemm-Str. 9, D-48149 M\"unster, Germany},
        C. Gebert\addressmark[DESY]\thanks{Talk given by Claus Gebert.},
        I. Montvay\addressmark[DESY],
        W. Schroers\address{Fachbereich Physik, Bergische Universit\"at, 
        Gesamthochschule Wuppertal, Gau\ss str. 20, \\ D-42097 Wuppertal, 
        Germany}}
       
\begin{document}

\begin{abstract}
We report about a simulation using three dynamical Wilson quarks and on the
progress in going to small quark masses.
\vspace{1pc}
\end{abstract}

\maketitle

\section{INTRODUCTION}

One of the main purposes of lattice QCD simulations is to predict
the spectrum of QCD, i.e. masses and decay constants of the 
light hadrons.
Due to computational and algorithmic limitations certain simplifications
have to be made. The quenched approximation allows to reproduce the physical 
spectrum already up to systematic errors of 10\%, as it was shown in detail
in \cite{Aoki99}.

Lots of effort has been spent to increase the available computing power, 
see e.g. \cite{Simm01} for such a project, and to overcome algorithmic 
limitations \cite{Mont96}. There have been several large
scale simulations of QCD with two dynamical, degenerate quarks \cite{Lipp01}, 
all based on the Hybrid Monte Carlo algorithm \cite{Duan87}. 
Still these simulations worked with relatively large quark masses.
For phenomenological reasons it is crucial to go beyond this scenario 
and simulate with three dynamical quarks with masses eventually approaching 
the physical ones \cite{Shar00}.

Some restrictions can be overcome
by using variants of the multibosonic algorithm \cite{Lusc94} like the
two-step multibosonic (TSMB) algorithm \cite{Mont96}. The latter has already 
been used successfully in simulations of supersymmetric field theories 
\cite{Camp99} and finite density QCD \cite{Hand00}.

In this contribution we present preliminary results from a first
simulation using three dynamical Wilson quarks. The aim of these simulations
is to achieve small quark masses and to estimate their influence on the 
spectrum.

\section{SMALL MASSES}

There are several problems when tuning the algorithms to small quark masses.
One problem is the critical slowing down, which reflects the decreasing 
efficency at smaller masses. In \cite{Schr01} it has been shown that this 
problem is less severe with multibosonic algorithms than with the HMC
algorithm. But still this can become a problem at too small masses.
A first orientation can be obtained by making a comparison with previous 
simulations. In \cite{Farc01} TSMB has been used for the $N=1$
supersymmetric Yang-Mills theory where the number of flavour is
$N_f=\frac{1}{2}$.
In that simulation condition numbers of ${\cal O}(10^5)$ occurred.
Because the performance of the algorithm depends mainly on the 
condition number $\lambda/\epsilon$ of the squared fermion matrix,
and not on $N_f$, one can similarly expect to reach condition numbers of at 
least ${\cal O}(10^5)$ for the case $N_f=3$. This would correspond to 
quark masses of 
roughly $\frac{1}{4}m_s$. The polynomial orders $n_1$ and $n_2$ that are 
needed for the TSMB algorithm mainly depend on $\lambda/\epsilon$. 
Changing from $N_f=\frac{1}{2}$ to $N_f=3$, $n_1$ is increasing by less than 
$50\%$ and $n_2$ by about $25\%$. This dependence is reflected by the 
asymptotic estimate of $n_2$ for large $\lambda/\epsilon$ \cite{Mont98}
\begin{equation}
n_2\simeq C_{N_f}\sqrt{\lambda/\epsilon}\propto(am_0)^{-1},
\end{equation}
where $C_{N_f}$ changes only slightly with $N_f$ and $m_0$ is the bare
quark mass.

Another potential problem when approaching the chiral limit may be posed by a 
sign change of the fermionic determinant. This so-called sign problem may 
spoil the statistical signal of observables. Experience shows that both
situations (with and without sign problem) are possible \cite{Scor01,Farc01}.
The sign of the determinant can be controlled by the spectral flow of small
eigenvalues. From this experience it is already clear that a sign change can 
only occur if there are extremly small eigenvalues (${\cal O}(10^{-8})$). Such 
eigenvalues did not appear in our runs so far, which gives
a first evidence for the absence of the sign problem.

\section{PARAMETER SPACE}

The spectrum of the light hadrons can be studied in the confined phase,
the phase below the finite temperature phase transition
\begin{equation}
T = (N_ta)^{-1} < T_c.
\end{equation}

\vspace{-20pt}

\begin{figure}[htb]
\psfig{file=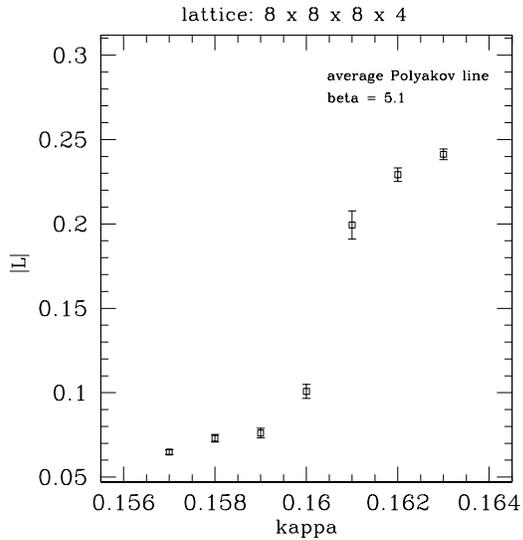,scale=.35}
\vspace{-25pt}

\caption{Finite temperature phase transition}
\label{fig:finT}
\end{figure}

\vspace{-14pt}

\noindent On small lattices the transition between the two phases can be 
seen clearly, 
as e.g. in figure~\ref{fig:finT}. The parameters of our simulations lead to
\begin{equation}
(T_cr_0)^{-1}\approx 2,
\end{equation}
where $r_0$ is the Sommer scale parameter.
Therefore we are safe from finite size effects due to deconfinement by
choosing lattice sizes $L$ such that
\begin{equation}
L/r_0 \ge 4.
\end{equation}

The parameters we finally want to achieve are shown in table~\ref{tab:para}.
These parameters give a constant volume $L/r_0=4$ and are such that
$(r_0m_{PS})^2=0.8$, corresponding to a quark mass of
$m_q\simeq 6.35m_{ud}\simeq 0.25m_s$. Since the volume and the quark mass
are constant one can perform a continuum extrapolation from these points.
In addition pion induced finite volume effects are under control as well, 
as they can be estimated by $\exp(-Lm_{PS})\simeq 0.03$.
\begin{table}[htb]
{
\vspace{-14pt}

$
\begin{array}{|c|c|c|c|c|}\hline
L/a& 0.5 a/r_0 & r_0/a& am_{PS}& am_q
\\ \hline\hline
8 & 0.250 & 2.0 & 0.44 & 0.032  \\ \hline
12& 0.167& 3.0 & 0.30 & 0.021  \\ \hline
16& 0.125& 4.0& 0.22& 0.016 \\ \hline
24& 0.083& 6.0 & 0.15 & 0.011  \\ \hline
\end{array}
$
\vspace{6pt}

$
\begin{array}{|c|c|c|c|}\hline
L/a&  \lambda/\epsilon & n_1 & n_2
\\ \hline\hline
8 &  1\cdot 10^4 & 48-56& 450 \\ \hline
12&  2\cdot 10^4 & 64-72& 600 \\ \hline
16&  4\cdot 10^4& 72-80& 900 \\ \hline
24&  9\cdot 10^4& 80-90 & 1200 \\ \hline
\end{array}
$
}
\vspace{5pt}

\caption{Estimates of parameters for the runs}
\label{tab:para}
\end{table}

\vspace{-15pt}

\begin{figure}[htb]
\vspace{9pt}
\psfig{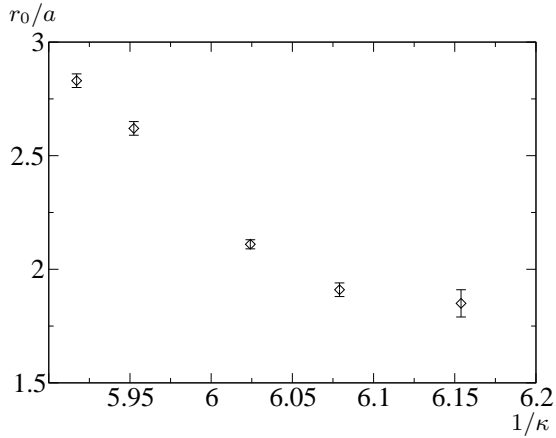}
\caption{Sommer scale $r_0$ for $\beta=5.1$}
\label{fig:r0}
\begin{picture}(0,0)
\put(0,195){\small\mbox{$r_0/a$}}
\put(190,40){\small\mbox{$1/\kappa$}}
\end{picture}
\vspace{-15pt}
\end{figure}

Results for our reference scale, the Sommer parameter $r_0$, are presented in 
figure~\ref{fig:r0}.

\begin{figure}[htb]
\vspace{9pt}
\hspace{-5pt}
\psfig{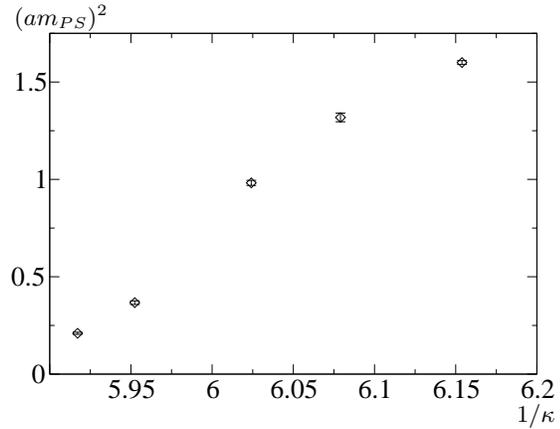}
\caption{Pseudoscalar mass $m_{PS}$ for $\beta=5.1$}
\label{fig:mps}
\begin{picture}(0,0)
\put(0,190){\small\mbox{$(am_{PS})^2$}}
\put(190,40){\small\mbox{$1/\kappa$}}
\end{picture}
\vspace{-20pt}
\end{figure}

\begin{figure}[htb]
\vspace{9pt}
\hspace{-5pt}
\psfig{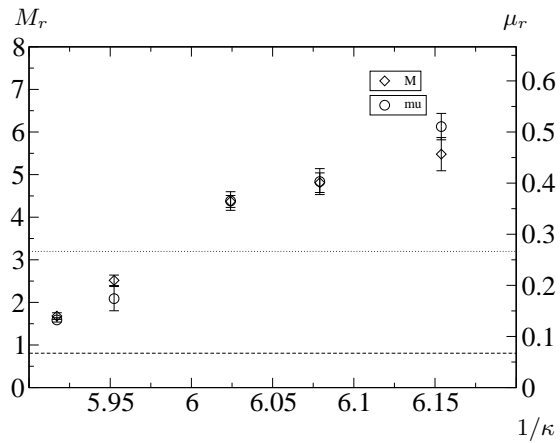}
\caption{Two definitions of the quark mass}
\label{fig:mq}
\begin{picture}(0,0)
\put(0,195){\small\mbox{$M_r$}}
\put(185,195){\small\mbox{$\mu_r$}}
\put(190,40){\small\mbox{$1/\kappa$}}
\end{picture}
\vspace{-15pt}
\end{figure}

For tuning the parameters to the values in table~\ref{tab:para}
we further have to look at the pseudoscalar mass $m_{PS}$ and the quark mass.
This latter can be characterized by the dimensionless quantities
\begin{equation}
\mu_r=r_0m_q=\frac{r_0}{2}\frac{F_{PS}}{G_{PS}}m_{PS}^2
\end{equation}
and
\begin{equation}
M_r = (r_0m_{PS})^2.
\end{equation}
The two definitions become proportional to each other for small
masses. First results are shown in figures~\ref{fig:mps} and \ref{fig:mq}.
In figure~\ref{fig:mq} there are two horizontal lines indicating $m_s$ and
the mass region we want to reach, $\frac{1}{4}m_s$. 
Depending on the size of the mass $8^3\times 16$, $12^3\times 24$ 
and $16^3\times 32$ lattices have been used.

\section{CONCLUSIONS AND OUTLOOK}

We presented preliminary results from simulations with three dynamical quarks. 
The results show that it should be possible to reach small quark masses
within the proposed settings. For the moment we are exploring the region 
below the strange quark mass.

\section*{ACKNOWLEDGMENTS}

W.S. is supported by the DFG Graduiertenkolleg
``Feldtheoretische Methoden in der Elemen\-tar\-teil\-chen\-theorie
und Statistischen Physik''. The numerical productions were run on the
APEmille systems installed at NIC Zeuthen, the Cray T3E systems at
NIC J\"ulich, the ALiCE-cluster at Wuppertal University and the PC clusters 
at DESY Hamburg.



\begin{thebibliography}{9}
\bibitem{Aoki99} S. Aoki {\em et al.} [CP-PACS Collaboration],
                    Phys. Rev. Lett. {\bf 84} (2000) 238-241.
\bibitem{Simm01} H. Simma {\em et al.} [APE Collaboration],
                    these proceedings.
\bibitem{Mont96} I. Montvay, Nucl. Phys. {\bf B466} (1996) 259-284.
\bibitem{Lipp01} Th. Lippert, these proceedings; A. Ukawa, ibid;
                 H. Wittig, ibid.
\bibitem{Duan87} S. Duane, A. D. Kennedy, B. J. Pendleton, D. Roweth,
                    Phys. Lett. {\bf B195} (1987) 216-222.
\bibitem{Shar00} S. Sharpe, N. Shoresh, Phys. Rev. {\bf D62} (2000) 094503. 
\bibitem{Lusc94} M. L\"uscher, Nucl. Phys. {\bf B418} (1994) 637-648.
\bibitem{Camp99} I. Campos {\em et al.} [DESY-M\"unster Collaboration],
                    Eur. Phys. J. {\bf C11} (1999) 507-527.
\bibitem{Hand00} S. Hands {\em et al.}, Eur. Phys. J. {\bf C17}
                    (2000) 285-302.
\bibitem{Schr01} W. Schroers {\em et al.}, hep-lat/0110033.
\bibitem{Farc01} F. Farchioni {\em et al.}, these proceedings.
\bibitem{Mont98} I. Montvay, Comput. Phys. Commun. {\bf 109} (1998) 144-160. 
\bibitem{Scor01} L. Scorzato {\em et al.}, these proceedings.
\end{thebibliography}
\end{document}